\documentclass[aps, nofootinbib,superscriptaddress, preprintnumbers, longbibliography]{article}

\usepackage{jheppub}
\usepackage{amsmath,amsfonts,amssymb}
\usepackage{bbm}
\usepackage{hyperref}
\usepackage{cleveref}
\usepackage{color}
\usepackage{comment}
\usepackage{caption}
\usepackage[latin9]{inputenc}
\usepackage{array}
\usepackage{booktabs}
\usepackage{breakurl}
\usepackage{subcaption}
\usepackage{ragged2e}
\usepackage{tikz}

\usepackage{graphicx}
\usepackage{esint}

\newcommand{\mFP}{m_\mathrm{FP}}
\newcommand{\gmn}{g_{\mu\nu}}
\newcommand{\fmn}{f_{\mu\nu}}
\newcommand{\dd}{\mathrm{d}}

\begin{document}

\title{Higuchi Bound on Slow Roll Inflation and the Swampland}

\author[a]{Marvin L\"uben,}
\author[a,b]{Dieter L\"ust}

\affiliation[a]{Max-Planck-Institut f\"ur Physik (Werner-Heisenberg-Institut),\\
F\"ohringer Ring 6, 80805 Munich, Germany}
\affiliation[b]{Arnold-Sommerfeld-Center for Theoretical Physics, Ludwig-Maximilians-Universit\"at, \\
Theresienstr. 37, 80333 M\"unchen, Germany}

\emailAdd{mlueben@mpp.mpg.de}
\emailAdd{dieter.luest@lmu.de}

\begin{flushright} 
\texttt{LMU-ASC 10/20    MPP-2020-31}
\end{flushright}

\abstract{
In this paper we study the implications of the generalized Higuchi bound on massive
spin-two fields for the derivative of the scalar potential within bimetric theory.
In contrast to the recent de Sitter swampland conjecture, an upper bound 
on the derivate of the scalar potential follows from the generalized Higuchi bound.
In combination, this leaves a window for the derivate of the scalar potential.
We discuss this bound in several representative bimetric models and parameter regions.}

\maketitle

\section{Introduction}

As it is extensively discussed during the recent two years, the effective potential 
of scalar fields in effective gravitational theories is severely constrained by fundamental quantum gravity or string theory
considerations 
\cite{Dvali:2013eja,Obied:2018sgi,Dvali:2018fqu,Ooguri:2018wrx,Dvali:2018jhn,Palti:2019pca}. These constraints have profound implications on the evolution of the universe, in particular on inflationary scenarios and
on the problem of dark energy \cite{Agrawal:2018own,Garg:2018reu}.

A genuine feature of quantum gravity and string theory is the
existence of higher spin fields.
The effective field theory of higher
spin fields is highly constrained.
One such constraint is the Higuchi bound~\cite{Higuchi:1986py},
\begin{flalign}
	m^2\geq s(s-1)H^2
\end{flalign}
for a massive (bosonic) field of spin $s$ and mass $m$
that propagates in de Sitter (dS) spacetime with radius $1/H$
(with $H$ the Hubble parameter).
If the bound is violated the massive higher spin field contains
helicity modes with negative norm which is in conflict with
unitarity.
Recently some implications from the Higuchi bound on massive higher spin
states in string theory and on the scale of inflation were 
discussed in~\cite{Lust:2019lmq,Noumi:2019ohm,Scalisi:2019gfv}.

The Higuchi bound is non-trivial for $s\geq2$.
In this article we focus on massive spin-2 fields~\cite{Pauli:1939xp,Fierz:1939ix}
with Fierz-Pauli mass $\mFP$.
Here the Higuchi bound explicitly reads $\mFP^2 \geq 2H^2$.
In the case where the Hubble parameter is given by the potential energy
of a scalar field, i.e. $3H^2=V$, the Higuchi bound provides
an upper limit on the potential energy as
\begin{flalign}
	3\mFP^2\geq2V
\end{flalign}
given the mass of the spin-2 field.

In Ref.~\cite{Fasiello:2013woa}
the Higuchi bound of dS spacetime was generalized for FLRW spacetime
to a cosmological stability bound.
This bound was found in the context of bimetric
theory~\cite{Hassan:2011zd,Hassan:2011ea} 
which serves as the low-energy effective field theory for
a massive and a massless spin-2 field
with fully non-linear (self-)interactions.
In~\cref{sec:ConstraintOnDerivativeScalarPot} we examine
implications of this bound in bimetric theory for the
derivative of the scalar potential, denoted by $V'$.
In contrast to the swampland condition of~\cite{Obied:2018sgi}, where
a lower bound on $V'$ is conjectured,
we will show that an upper bound on $V'$ follows
from the generalized Higuchi bound.

\section{Ingredients from bimetric theory}

In order to describe a massive spin-2 field propagating on
a curved background such as FLRW, the linear theory by Fierz and
Pauli~\cite{Pauli:1939xp,Fierz:1939ix,Vainshtein:1972sx,Boulware:1973my}
must be completed by
non-linear terms~\cite{Bernard:2014bfa,Bernard:2015mkk,Bernard:2015uic}.
Bimetric theory is the ghost-free non-linear completion of Fierz-Pauli
describing a gravitating massive spin-2 field
~\cite{deRham:2010ik,deRham:2010kj,Hassan:2011zd,Hassan:2011ea}.
For a review we refer to Ref.~\cite{Schmidt-May:2015vnx}.
Furthermore, for some work about bimetric theory in relation to
the swampland and string theory realizations
see~\cite{Bachas:2018zmb,Ferrara:2018wlb,Klaewer:2018yxi,deRham:2018dqm,Bachas:2019rfq}.

\subsection{Action and equations of motion}

We focus on singly-coupled bimetric theory where the Standard Model couples to only
one of the metric tensors, say $\gmn$.
The action for the metric tensors $\gmn$ and $\fmn$ is given
by~\cite{Hassan:2011zd}
\begin{flalign}\label{eq:bimetric-action}
	S & = \frac{M_p^2}{2}\int\mathrm{d}^4x\left(\sqrt{-g} R(g)+ \alpha^2\sqrt{-f}R(f)\right)\\
	&- M_p^2\int\mathrm{d}^4x\sqrt{-g}\, \mathcal U(g,f)
	+ \int\mathrm{d}^4x\sqrt{-g}\mathcal L_\mathrm{m}(\gmn,\Phi)\nonumber\,.
\end{flalign}
where $R(g)$ and $R(f)$ are the Ricci scalars of $\gmn$ and $\fmn$, resp.
The quantity $M_p$ is the Planck mass of $\gmn$ and $\alpha=M_f/M_p$ measures its
ratio to the Planck mass of $\fmn$.
The bimetric potential $\mathcal U(g,f)$ is given in
terms of the square-root matrix $S^\mu_{\ \nu}$ defined by the relation
\begin{flalign}\label{eq:square-root-matrix}
	S^\mu_{\ \alpha}S^\alpha_{\ \nu}=g^{\mu\alpha}f_{\alpha\nu}\,,
\end{flalign}
which in general does not have an unique solution.
Only if the metrics $g_{\mu\nu}$ and $f_{\mu\nu}$
share the same time- and space-like directions,
the theory is defined unambiguously and an unique and real solution to~\cref{eq:square-root-matrix}
exists~\cite{Hassan:2017ugh}.
Then, the bimetric potential reads
\begin{flalign}
	\mathcal U(g,f)=\sum_{n=0}^4\beta_n e_n(S)
\end{flalign}
in terms of the elementary symmetric polynomials $e_n$.
The parameters $\beta_n$ are real and have mass dimension $2$ in this parametrization.
Note that $\beta_0$ and $\beta_4$ parametrize the  vacuum energy in the
$\gmn$- and $\fmn$-sector, resp.
Finally, Standard Model fields are collectively denoted as $\Phi$.

Varying the action~(\ref{eq:bimetric-action}) with respect to $g^{\mu\nu}$ and $f^{\mu\nu}$
yields two sets of modified Einstein equations,
\begin{flalign}\label{eq:Einstein}
	G_{\mu\nu}^g+\mathcal U^g_{\mu\nu}=\frac{1}{M_p^2}T_{\mu\nu}\,,\quad
	\alpha^2G_{\mu\nu}^f+\mathcal U^f_{\mu\nu}=0\,,
\end{flalign}
where $G^g_{\mu\nu}$ and $G^f_{\mu\nu}$ are the Einstein tensors of $\gmn$ and $\fmn$,
resp. The explicit form of contributions coming from the bimetric potential 
$\mathcal U^g_{\mu\nu}$ and $\mathcal U^f_{\mu\nu}$ can be found, e.g.,
in Ref.~\cite{Hassan:2011zd,Hassan:2011vm}.
The stress-energy tensor of matter is defined as
\begin{flalign}
	T_{\mu\nu}=\frac{-2}{\sqrt{-g}}\frac{\partial \sqrt{-g}\mathcal L_\mathrm{m}}{\partial g^{\mu\nu}}\,.
\end{flalign}
If the matter-sector $\sqrt{-g}\mathcal L_\mathrm{m}$ is invariant under diffeomorphisms,
stress-energy is conserved, $\nabla^\mu T_{\mu\nu}=0$, where $\nabla_\mu$ is
the covariant derivative compatible with $g_{\mu\nu}$.
The Bianchi identity $\nabla^\mu G^{g}_{\mu\nu}=0$ implies the 
so-called Bianchi constraint
\begin{flalign}
	\nabla^\mu \mathcal U^{g}_{\mu\nu}=0\,.
\end{flalign}

\subsection{Proportional background solutions}

An important class of solutions are the proportional backgrounds where
both metric are related by a conformal factor,
$\fmn=c^2\gmn$~\cite{Hassan:2012wr}.
The Bianchi constraint forces $c$ to be a constant.
These solutions are maximally symmetric, i.e. (A)dS or Minkowski.
The effective cosmological constant in terms of the bimetric parameters
is given by
\begin{flalign}\label{eq:Bimetric-Lambda}
	\Lambda& = V + \beta_0+3\beta_1 c+3\beta_2 c^2 + \beta_3 c^3\nonumber\\
	&=\frac{1}{\alpha^2 c^2}(\beta_1 c+3\beta_2 c^2 +3\beta_3 c^3 +\beta_4 c^4)\,.
\end{flalign}
The equality of the first and second line follows from the proportionality of the
Einstein tensors for $\gmn$ and $\fmn$.
Here, we already anticipated that the matter sector contributes potential energy $V$
to the effective cosmological constant.
Therefore $\Lambda$ is composed out of vacuum energy, interaction energy of the massive
spin-2 fields, and potential energy from the matter sector\footnote{Note
that the vacuum energy contribution $\beta_0$ is degenerate with an unspecified
potential $V$.
In the analysis of this paper however, $V$ serves as an input parameter while
the interaction parameters are fixed in terms of $V$, $\mFP$ and $c$.
Later we will go to the parameter regime where
$\Lambda\simeq V$ or equivalently $V\gg\beta_0+3\beta_1 c+3\beta_2 c^2+\beta_3 c^3$.
For these reasons we treat $V$ and $\beta_0$ as independent parameters.}.
Eq.~(\ref{eq:Bimetric-Lambda}) can be thought of as determining the
proportionality constant $c$. 

Bimetric theory has a well-defined mass spectrum around proportional
backgrounds.
The metric fluctuations are a linear superposition of mass eigenstates.
The mass of the massive spin-2 field is given by
\begin{flalign}\label{eq:FP-mass}
	\mFP^2=\frac{1+\alpha^2 c^2}{\alpha^2 c^2}c( \beta_1 + 2\beta_2 c+ \beta_3 c^2 )
\end{flalign}
in terms of the bimetric parameters. We refer to this quantity as Fierz-Pauli mass.
As mentioned previously, the Fierz-Pauli mass is subject to the
Higuchi bound~\cite{Higuchi:1986py},
\begin{flalign}
	3\mFP^2>2\Lambda
\end{flalign}
to ensure unitarity.

\subsection{Flat FLRW solutions}

Assuming homogeneity and isotropy according to the cosmological principle,
both metrics are of FLRW from~\cite{Volkov:2011an,vonStrauss:2011mq,Comelli:2011zm},
\begin{subequations}\label{eq:FLRW-ansatz}
\begin{flalign}
	\dd s_g^2=& -\dd t^2 + a^2 \dd \vec x^2\\
	\dd s_f^2=&- X^2\dd t^2 + b^2 \dd \vec x^2\,.
\end{flalign}
\end{subequations}
Here, the functions $a$ and $b$ are the scale factors of $\gmn$ and $\fmn$, resp.,
while $X$ is the lapse of $\fmn$.
All metric functions depend on time $t$ only.
We introduce the Hubble rate $H$ and the scale factor ratio $y$
as
\begin{flalign}
	H=\frac{\dot a}{a}\ , \ \ y=\frac{b}{a}\,,
\end{flalign}
where a dot denotes derivative with respect to cosmic time $t$.

Next, we assume that the matter sector consists of a real
scalar field $\phi$ with Lagrangian
\begin{flalign}
	\mathcal L_{\rm m}=-\frac{1}{2}g^{\mu\nu}\partial_\mu\phi\partial_\nu\phi-V(\phi)
\end{flalign}
with scalar potential $V(\phi)$.
According to the cosmological principle, the scalar field can only
depend on time $t$.
Consequently, the energy density and pressure are given by
\begin{flalign}\label{eq:scalar-energy-density}
	\frac{\rho_\phi}{M_p^2}=\frac{1}{2}\dot\phi^2+V\ , \ \ \frac{p_\phi}{M_p^2}=\frac{1}{2}\dot\phi^2-V
\end{flalign}
The stress-energy tensor is given by
\begin{flalign}
	T^\mu_{\ \nu}={\rm diag}(-\rho_\phi,p_\phi,p_\phi,p_\phi)\,.
\end{flalign}
The equation of motion for the scalar field reads
\begin{flalign}\label{eq:scalar-eom}
	\ddot\phi+3H\dot\phi+V_\phi=0\,.
\end{flalign}

Next, we derive the Friedmann equations.
The Bianchi constraint is solved by
\begin{flalign}
	X\dot a = \dot b\,,
\end{flalign}
on the so-called dynamical branch\footnote{In
principle, another solution to the Bianchi constraint exists.
On this algebraic branch, $y$ is forced to be a constant
and the solution has various pathologies~\cite{Comelli:2012db,Cusin:2015tmf}.}.
Plugging the ansatz~(\ref{eq:FLRW-ansatz}) into the modified
Einstein~\cref{eq:Einstein} yields two sets of Friedmann equations.
These can be rearranged to give a single Friedmann
equation.
Here, we adopt the notation of Ref.~\cite{Sakakihara:2016ubu}
where the Friedmann equation reads
\begin{flalign}\label{eq:simple-Friedmann}
	3H^2=\frac{\rho_\phi}{M_{\rm eff}^2}
\end{flalign}
in terms of the time-varying effective Planck mass
\begin{flalign}
	M_{\rm eff}^2 = \frac{1+\alpha^2 y^2}{1+\frac{U}{\rho_\phi}} M_p^2\,.
\end{flalign}
The function $U$ depends on the parameters of the bimetric potential as
\begin{flalign}
	U=M_p^2\left(\beta_0+4\beta_1 y+6\beta_2 y^2+4\beta_3 y^3+\beta_4 y^4\right)\,.
\end{flalign}
The scale factor ratio is determined by a quartic polynomial,
\begin{flalign}\label{eq:quartic-pol}
	P(y)=\rho_\phi\,,
\end{flalign}
where the polynomial $P(y)$ is given by
\begin{flalign}
	P(y)= \frac{1+\alpha^2 y^2}{4\alpha^2 y} U'-U
\end{flalign}
in terms of $U$ with $U'=\dd U/\dd y$.
The consistency of the solutions to this equation have been studied in detail in
Refs.~\cite{Konnig:2013gxa,Konnig:2015lfa,Luben:2020xll}.
However, in this note we will adopt a perturbative treatment.

\subsection{Generalized Higuchi bound}

By inspecting the sign of the kinetic term of the scalar mode
in the minisuperspace approximation, the authors of
Ref.~\cite{Fasiello:2013woa} derived
a cosmological stability bound that generalizes the Higuchi bound
to FLRW spacetime.
Let us define a time dependent mass parameter as
\begin{flalign}\label{eq:gen-FP-mass}
	m_{\rm gen}^2=\frac{1+\alpha^2 y^2}{\alpha^2 y}(\beta_1+2\beta_2 y+\beta_3 y^2)\,.
\end{flalign}
This quantity reduces to the Fierz-Pauli mass~(\ref{eq:FP-mass}) when
the metrics are proportional\footnote{Note,
that a well-defined notion of mass in the representation-theoretic sense
does not exist in FLRW.
In order to define a mass, the background spacetime needs to satisfy
Lorentz- or (A)dS-isometries.}, i.e., for $y=X=c$.
It was shown that the kinetic term of the scalar mode is multiplied by
\begin{flalign}
	W=m_{\rm gen}^2-2H^2\,.
\end{flalign}
The kinetic term of the scalar mode has the correct sign if
\begin{flalign}
	W > 0\,.
\end{flalign}
This is the generalized Higuchi bound that we will use in the next section.
Note, that this bound reduces to the standard Higuchi bound when the metrics
are proportional.

\section{Constraint on the derivative of the scalar potential}
\label{sec:ConstraintOnDerivativeScalarPot}

As mentioned previously, the Higuchi bound provides an upper limit on the
(local) minimum of the scalar potential.
By studying small displacements from the minimum, we now derive a
bound on the derivative of the potential close to its minimum.
We will work in the slow-roll approximation.
We assume that $\Lambda\simeq V$, i.e., the potential energy
of the scalar field dominates over the vacuum and interaction
energy.
We further assume that the Higuchi bound is not saturated, i.e.,
we exclude the partially-massless case~\cite{Deser:1983mm,Hassan:2012gz},
in order to avoid devision by zero.

First, we express the energy density of the scalar field
in terms of the potential $V$ and its derivative.
Let us introduce the slow-roll parameter
\begin{flalign}
	\epsilon=\frac{M^2_{\rm eff}}{2}\left(\frac{V'}{V}\right)^2\,,
\end{flalign}
where $V'$ means derivative of $V$ with respect to the scalar field $\phi$.
In the slow-roll approximation $\epsilon\ll1$ and $\ddot\phi\ll H\dot\phi$, the
scalar equation of motion~(\ref{eq:scalar-eom}) can be rearranged to
\begin{flalign}
	\dot\phi=-\frac{V'}{3H}=-\frac{\sqrt{2\epsilon}V}{3M_{\rm eff} H}\,.
\end{flalign}
Replacing $H$ using the Friedmann~\cref{eq:simple-Friedmann,eq:scalar-energy-density}
yields after rearranging
\begin{flalign}
	\left(\frac{1}{2}\dot\phi^2+V\right)\dot\phi^2=\frac{2\epsilon}{3}V^2\,.
\end{flalign}
Solving for $\dot\phi$ yields $\dot\phi^2=-V(1\pm\sqrt{1+4\epsilon/3})$ out of which
only the '$-$' branch is physical (as only then $\dot\phi=0$ for $\epsilon=0$).
Plugging this back into our expression for the energy density yields
\begin{flalign}\label{eq:energy-density-sr}
	\frac{\rho_\phi}{M_p^2}=V\left(1+\frac{\epsilon}{3}\right)
\end{flalign}
up to first order in slow-roll.
Plugging this into the Friedmann equation yields
\begin{flalign}
	3H^2=\frac{\rho_\phi}{M_{\rm eff}^2}=V\left(1+\frac{\epsilon}{3}\right)\,.
\end{flalign}
Here we used that $M_{\rm eff}/M_p\sim\mathcal O(1)$ during slow-roll.

To find an expression for the generalized spin-2 mass $m_{\rm gen}$ up to first order
in slow-roll we plug~\cref{eq:energy-density-sr} into the quartic
polynomial~(\ref{eq:quartic-pol}) and solve for $y$ order by order in $\epsilon$.
This results in
\begin{flalign}
	y=c \left( 1-\frac{V}{3\mFP^2-2V} \frac{\epsilon}{3} \right)\,.
\end{flalign}
Here, $c$ is the solution for $\epsilon=0$, i.e., at the de Sitter point where the cosmological constant
is given by the potential energy of the scalar field only. The Fierz-Pauli mass
is defined on that background as in~\cref{eq:FP-mass}.
Plugging this into~\cref{eq:gen-FP-mass} yields
\begin{flalign}
	m_{\rm gen}^2=\mFP^2 + F\, \frac{\epsilon}{3}
\end{flalign}
up to first order in slow-roll.
The quantity $F$ depends on the bimetric parameters and scalar potential as
\begin{flalign}
	F=-\frac{V c }{3\mFP^2-2V}\left(\frac{\beta_3c^2-\beta_1}{\alpha^2 c^2}+\beta_1+4\beta_2 c+3\beta_3c^2\right)\,.\nonumber
\end{flalign}
The expression in parenthesis does not combine into well-defined background
quantities.

Combining our ingredients, we find $W$ in the slow-roll approximation as
\begin{flalign}
	0\leq W=3\mFP^2-2V + \left(F-\frac{2}{3}V\right)\epsilon\,.
\end{flalign}
For $F\ge 2V/3$ the bound on $\epsilon$ is trivial.
However, if the Higuchi bound is satisfied and if $F<2V/3$, we find the following non-trivial bound on $\epsilon$:
\begin{flalign}\label{eq:bound-on-e}
	\epsilon \leq 3 \frac{3\mFP^2-2V}{2V-3F}\,.
\end{flalign}
In~\cref{sec:explicit-F} we discuss typical values of the right hand side
of~\cref{eq:bound-on-e} by studying various simplified models
and different regions of the parameter space.
We generally find that the right hand side is of the order
of $\mFP^2/V$.
That means, if the mass of the spin-2 field is close to the Higuchi bound,
the restriction on $V'$ is most stringent as we assumed slow-roll to start with.

Finally, the bound~(\ref{eq:bound-on-e}) implies for the derivative of the scalar potential,
\begin{flalign}
	\frac{| V' |}{V} \lesssim \frac{\sqrt{6}}{M_p} \sqrt{\frac{3\mFP^2-2V}{2V-3F}}\,.
\end{flalign}
Here we used again that generally $M_{\rm eff}/M_p\sim\mathcal O (1)$.
Hence, we found an upper limit on the absolute value of the derivative of the scalar potential.
The restriction is most stringent close to the Higuchi bound.
Saturation of the bound is excluded and hence the right hand
side cannot vanish.

\section{Summary and discussion}

We used bimetric theory as the low-energy effective field theory
to describe a massive spin-2 field with mass $\mFP$ propagating on FLRW spacetime.
Assuming that the matter sector is composed out of a scalar field with potential $V$,
we showed that the generalized Higuchi bound implies 
an upper bound on the  parameter of the following form:
\begin{flalign}
	\epsilon \leq c\,\frac{\mFP^2}{V}\ , \quad {\rm for}\ c\sim \mathcal O(1)\,.
\end{flalign}
For the derivative of the scalar
potential an equivalent upper bound can be written as
\begin{flalign}\label{finalbound}
	\frac{| V' |}{V} \leq \frac{c}{M_p} \sqrt{\frac{\mFP^2}{V}}\,.
\end{flalign}
In our derivation we assumed that only the potential energy of the scalar field
contributes to an effective cosmological constant.
For a large spin-2 mass $\mFP^2\gg V$, this bound is not restrictive as we
assumed slow-roll (i.e. $\epsilon\ll1$) in our derivation.
On the other hand, if the mass of the spin-2 field is close to the Higuchi bound,
the derivative of the scalar potential is bounded from above.

Let us compare our result with the bounds on the scalar potential,
which are obtained in the context of the de Sitter swampland conjecture
\cite{Obied:2018sgi,Ooguri:2018wrx}, 
stating that 
\begin{equation}\label{swampland}
\frac {| V' |}{V}\geq {c'\over M_p} \qquad {\rm or}\qquad V''\leq -{c''\over M_p^2}V\, ,
\end{equation}
with $c',c''$ being some universal constants of order one.
This bound rules out de Sitter vacua and restricts
 inflationary scenarios in string theory.

Applying the lower bound from the de Sitter swampland conjecture and
combining it with the upper bound from the generalized Higuchi bound in
bimetric theory, it follows that 
the derivative of the scalar potential has to lie in the following window:
\begin{equation}\label{combinded swampland}
 {c'\over M_p}    \leq  \frac {| V' |}{V}  \leq \frac{c}{M_p} \sqrt{\frac{\mFP^2}{V}} \, .
\end{equation}
The window is small when the spin-2 mass is close to the Higuchi bound, in
which case the derivative has to be of the order $|V'|/V\simeq 1/M_p$.
However, due to the Higuchi bound, the window does not close completely. 
Let us point out some caveats again.
The Higuchi bound on $V'$ was found within the slow-roll approximation.
Further, there are specific parameter regions
where $V'$ is not bounded from above.
However, if the bound in~\cref{combinded swampland} on $V'$ is respected
the generalized Higuchi bound is guaranteed to be satisfied for generic parameter values.

In string theory  it is natural to assume that the spin-2 mass $\mFP$ is related
to higher spin string excitations~\cite{Ferrara:2018wlb}.
Assuming that~\cref{finalbound} holds generally
and identifying the spin-2 mass with the string scale, $\mFP\simeq M_s$, 
our bound reads
\begin{flalign}\label{finalbound1}
	\frac{| V' |}{V} \leq c\frac{M_s}{M_p} \sqrt{\frac{1}{V}}\simeq g_s \sqrt{\frac{1}{V{\cal V}}}\,.
\end{flalign}
Here $g_s$ is the string coupling constant, and ${\cal V}$ denotes
the internal volume of the compact six-dimensional space.
We can use the bound on $M_s$ that was obtained in Ref.~\cite{Lust:2019lmq},
\begin{equation}
M_s^2\geq \sqrt{\frac{V}{3}}M_p\,,
\end{equation}
which follows when one requires that the entire string Regge trajectory
satisfied the Higuchi bound.
So taking also this bound into account, the condition (\ref{finalbound1}) is
always satisfied if the following stronger bound holds:
\begin{flalign}\label{finalbound2}
	\frac{| V' |}{V} \leq \frac{c}{M_p} \sqrt{\frac{M_p}{(V/3)^{1/2}}}\,.
\end{flalign}

As a next step it would be interesting to study whether the upper bound on $V'$ is a generic implication
of unitarity when massive spin-2 fields propagate on dS spacetime.
This requires a treatment beyond slow-roll.
Further, it would be interesting to check, wether this combined bound is satisfied, e.g.,
in concrete string theory settings.

\vspace{5pt}
\subsubsection*{Acknowledgments} We are grateful to E. Palti for useful discussions.
This work is supported by a grant from the Max-Planck-Society. The work of D.L. is supported by the Origins Excellence Cluster.

\appendix

\section{Model-specific considerations}\label{sec:explicit-F}
In this section, we provide explicit expressions for the upper bound on $\epsilon$ in terms
of the background variables $\mFP$ and $V$ for several representative bimetric models
and parameter regions.
The relation between the interaction parameters $\beta_n$ and the background
quantities is determined by~\cref{eq:Bimetric-Lambda,eq:FP-mass}
as described in detail in Ref.~\cite{Luben:2020xll}.
Note that in general, the bimetric potential gives rise to an effective
cosmological constant $\Lambda_{\rm bim}=\beta_0+3\beta_1 c+3\beta_2 c^2+\beta_3 c^3$.
Here, we assume that only the scalar potential
contributes to the overall cosmological constant, i.e. we assume $\Lambda \simeq V$
and the contributions from vacuum energy and interaction energy are subdominant, i.e.,
$V\gg \Lambda_{\rm bim}$.

In the following, we only state the results and point out some caveats about the
more restricted submodels.
For brevity, let us define the maximum value of $\epsilon$ as
\begin{flalign}\label{eq:def-e-max}
	\epsilon_{\rm max} = \frac{3(3\mFP^2-2V)}{2V-3F}\,,
\end{flalign}
in terms of which the generalized Higuchi bound in slow-roll approximation
reads $\epsilon<\epsilon_{\rm max}$.

Before studying individual models we point out again that an unspecified scalar potential $V$
is degenerate with the vacuum energy parameter $\beta_0$.
In our analysis however, the potential $V$ is not unspecified.
Instead, from the Higuchi bound we derive an upper limit on $|V'|$ given
the values of $\mFP$ and $V$.
The interaction parameters $\beta_n$, including $\beta_0$,
are determined by~\cref{eq:Bimetric-Lambda,eq:FP-mass} in terms of $\mFP$ and $V$.

Setting $\beta_2=\beta_3=\beta_4=0$ in~\cref{eq:Bimetric-Lambda,eq:FP-mass},
solving for $\beta_0$ and $\beta_1$,
and plugging the results into~\cref{eq:def-e-max}
we find
\begin{flalign}
	\epsilon_{\rm max}^{01}=\frac{3(3\mFP^2-2V)^2}{(3\mFP^2-4V)V}
\end{flalign}
as the upper limit on $\epsilon$ in the $\beta_0\beta_1$-model\footnote{Here
we keep $\beta_0$ as a free parameter.
Setting $\beta_0$ to zero or absorbing the parameter into $V$ fixes the spin-2
mass to be $\mFP^2=V$.
Keeping $\beta_0$ and $V$ independent allows $\mFP$ to be independent.}.
This model is only well-defined for $\mFP^2>V$ which is
more restrictive than the Higuchi bound~\cite{Luben:2020xll}.
Spin-2 masses in the range $V<\mFP^2<\frac{4}{3}V$ imply
$\epsilon_{\rm max}^{01}<0$.
Since $\epsilon$ is a manifestly positive quantity,
the bound is trivial in this parameter region.
Expanding around $\mFP^2\gg V$ the expression simplifies to
$\epsilon_{\rm max}^{01}\simeq 9\mFP^2/V$.
Next, for the $\beta_1\beta_4$-model, we set $\beta_0=\beta_2=\beta_3=0$,
solve~\cref{eq:Bimetric-Lambda,eq:FP-mass} for $\beta_1$ and $\beta_4$
and plug the result into~\cref{eq:def-e-max}.
This leads to the same expression as for the $\beta_0\beta_1$-model,
$\epsilon_{\rm max}^{14}=\epsilon_{\rm max}^{01}$.
Therefore, close to the Higuchi bound the restriction on $\epsilon$ is most stringent.
Moving to the $\beta_1\beta_2$-model by setting $\beta_0=\beta_3=\beta_4=0$
the same procedure leads to
\begin{flalign}
	\epsilon_{\rm max}^{12} = \frac{3(3 \mFP^2-V)^2}{(15\mFP^2-7V)V}\,.
\end{flalign}
In the limit of large spin-2 mass we can approximate the bound as
$\epsilon^{12}_{\rm max}\simeq 9\mFP^2/(5V)$.
Finally for the $\beta_1\beta_3$-model we set $\beta_0=\beta_2=\beta_4=0$
to find
\begin{flalign}
	\epsilon^{13}_{\rm max} = \frac{3(3\mFP^2-2V)^2}{16(3\mFP^2-V)\mFP^2}\,.
\end{flalign}
Expanding for $\mFP^2\gg V$ this expression reduces to $\epsilon_{\rm max}^{13}\simeq 9/16$.
Therefore, the upper limit on $\epsilon$ in the $\beta_1\beta_3$-model differs significantly from
the upper limit in the other two parameter models.

Summarizing, for the two parameter models
we find that $\epsilon_{\rm max} \simeq \mFP^2/V$ in general.
Only the $\beta_1\beta_3$-model is an exception, which implies the stronger
bound $\epsilon_{\rm max}\leq 9/16$.

\begin{figure}
\centering
	\includegraphics[scale=0.59]{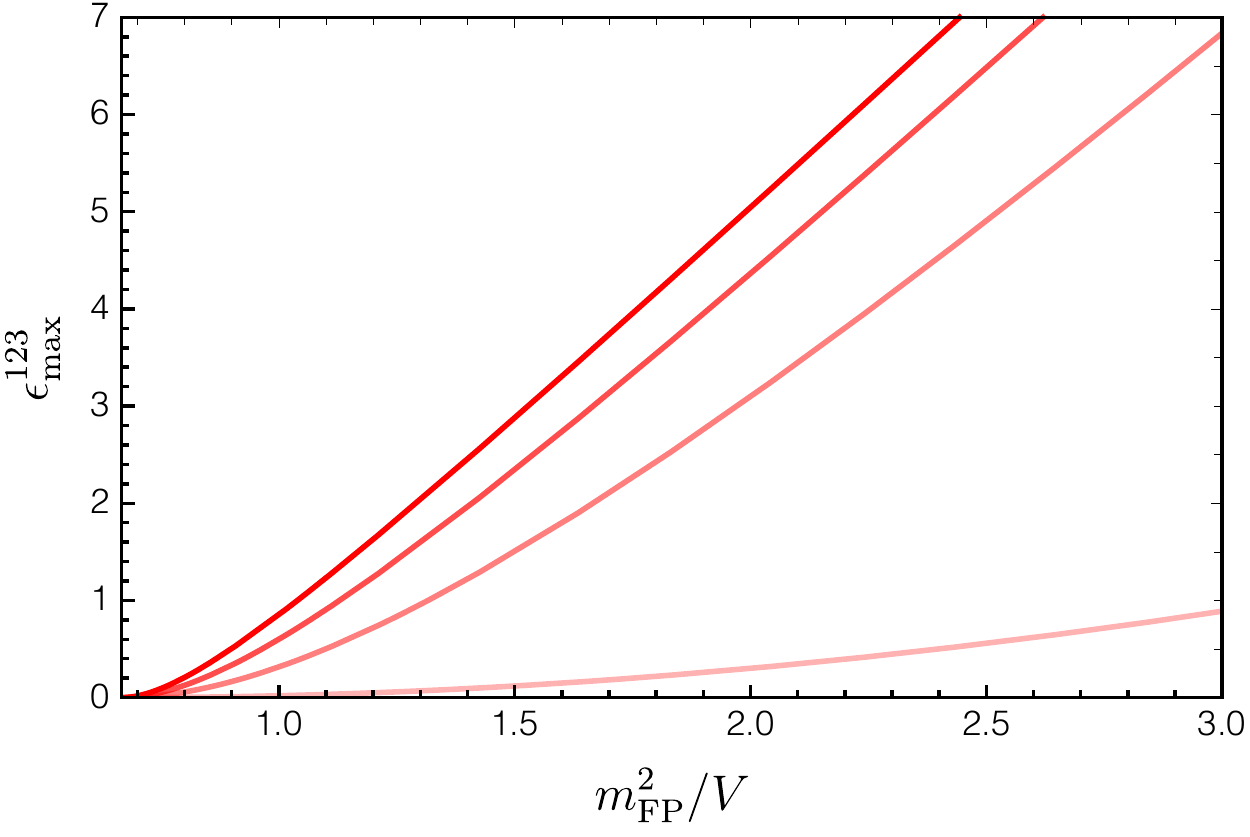}
	\caption{The upper limit on $\epsilon$ is shown
	as a function of $\mFP^2/V$ for the $\beta_1\beta_2\beta_3$-model 
	for different values of $\bar\alpha$.
	We chose the values $\bar\alpha=0.6$, $0.8$, $2$, $10$, where the lightest color
	corresponds to the largest value of $\bar\alpha$.}
	\label{fig:e-max}
\end{figure}

Next, we study more general cases by moving to three parameter models.
We focus on the $\beta_0\beta_1\beta_4$-model with vacuum energy for $\gmn$ and for $\fmn$
and on the $\beta_1\beta_2\beta_3$-model without vacuum energy.
For convenience, we introduce the quantity $\bar\alpha=\alpha c$~\cite{Luben:2020xll}.
For the $\beta_0\beta_1\beta_4$-model we set $\beta_2=\beta_3=0$,
solve~\cref{eq:Bimetric-Lambda,eq:FP-mass} for $\beta_0$, $\beta_1$ and $\beta_4$
and plug the result into~\cref{eq:def-e-max}.
This leads to the following upper limit on $\epsilon$,
\begin{flalign}
	\epsilon^{014}_{\rm max} = \frac{3(3\mFP^2-2V)^2}{(3\mFP^2-4V)V}\,,
\end{flalign}
which coincides with the expression that we obtained for the $\beta_0\beta_1$- and
$\beta_1\beta_4$-models.
Moving to the $\beta_1\beta_2\beta_3$-model by setting $\beta_0=\beta_4=0$,
the same procedure leads to
\begin{flalign}\label{eq:emax-123}
	\epsilon^{123}_{\rm max} = \frac{6(3\mFP^2-2V)^2}{(12\mFP^2-(5-3\bar\alpha^2)V)V}\,.
\end{flalign}
In~\cref{fig:e-max} we plot $\epsilon^{123}_{\rm max}$ as a function of $\mFP^2/V$
for different values of $\bar\alpha$.
For $\bar\alpha\lesssim 1$, the expression for $\epsilon^{123}_{\rm max}$ becomes independent
of $\bar\alpha$ and can be approximated by $\mFP^2/V$.
For larger values of $\bar\alpha$, the upper limit on $\epsilon$ is suppressed by
inverse powers of $\bar\alpha$.
This can be understood from expanding~\cref{eq:emax-123} in different limits.
For $\mFP^2\gg V$ the expression reduces to
$\epsilon^{123}_{\rm max}\simeq 9\mFP^2/(2V)$ which is independent of $\bar\alpha$.
Expanding for $\bar\alpha\gg1$ we find
$\epsilon^{123}_{\rm max}\simeq 2(9\mFP^4/V^2-12\mFP^2/V+4)/\bar\alpha^2$.
Therefore, in this limit $\epsilon^{123}_{\rm max}$ is suppressed by inverse powers of
$\bar\alpha$ and the leading term is proportional to $(\mFP^2/V)^2$
in contrast to our previously studied scenarios.
This parameter region however is excluded by consistency
conditions~\cite{Luben:2020xll}.
In the limit $\bar\alpha\ll1$, the upper limit reads
$\epsilon^{123}_{\rm max}\simeq 6(3\mFP^2-2V)^2/((12\mFP^2-5V)V)$
which is again independent of $\bar\alpha$.

Summarizing the results for the three parameter models, within the consistent parameter region
we find the same order of magnitude for the upper limit on $\epsilon$
as for the two parameter models.
That is $\epsilon_{\rm max}\sim \mFP^2/V$.
Only the $\beta_1\beta_2\beta_3$-model in the limit
$\bar\alpha\gg1$ is an exception.
Instead of discussing also the other three parameter models individually,
we now move to the full bimetric model with all interaction parameters left free.

With all interaction parameters left free, we
solve the background~\cref{eq:Bimetric-Lambda,eq:FP-mass}
for the interaction parameters
$\beta_1$, $\beta_2$, and $\beta_3$ in terms of the physical parameters
$\mFP$, $\Lambda=V$, and $\bar\alpha=\alpha c$. We also introduce
the quantity $\bar\beta_4=\alpha^{-4}\beta_4$.
In terms of these parameters, the upper limit on $\epsilon$ reads
\begin{flalign}
	\epsilon_{\rm max}=
	\frac{
	6\bar\alpha^2(3\mFP^2-2V)^2
	}{
	(3\beta_0
	+\bar\alpha^2(12\mFP^2-5V+3\beta_0)
	+3\bar\alpha^4(V-\bar\beta_4)
	-3\bar\alpha^6\bar\beta_4)V}\,.
\end{flalign}
Although this expression is quite lengthy, we can discuss the order of magnitude for
different parameter regimes.
In the limit of large spin-2 mass, $\mFP^2\gg V$, while keeping $\bar\alpha$ finite,
the expression reduces to
\begin{flalign}
	\epsilon_{\rm max}\simeq\frac{9}{2} \frac{\mFP^2}{V}\ , \ \ {\rm for}\ \ \mFP^2\gg V\,.
\end{flalign}
which is independent of $\bar\alpha$.
Next, we expand $\epsilon_{\rm max}$ for small (GR-limit) and for large (massive gravity limit)
$\bar\alpha$ resulting in
\begin{flalign}
	&\epsilon_{\rm max} = \frac{2\bar\alpha^2(3\mFP^2-2V)^2}{\beta_0 V}\ , \ \ {\rm for}\ \bar\alpha\ll1\,\nonumber\\
	&\epsilon_{\rm max} = -\frac{2(3\mFP^2-2V)^2}{\bar\alpha^4\, \bar\beta_4 V}\ , \ \ {\rm for}\ \bar\alpha\gg1\nonumber\,.
\end{flalign}
In both limits, the order of magnitude is set by $\mFP^4/(\beta_n V)$
and suppressed by (inverse) powers of $\bar\alpha$.
However, this conclusion is model-dependent. In our parametrization
the interaction parameters $\beta_0$ and $\bar\beta_4$ are fixed, while the other
parameters scale with $\bar\alpha$ in a nontrivial way.
We also studied the limit of small and large $\bar\alpha$ in other parametrizations.
We find that $\epsilon_{\rm max}$ is not necessarily suppressed by (inverse) powers
of $\bar\alpha$, while in these cases the order of magnitude is set by $\mFP^2/V$.

Summarizing, there are regions of the parameter space where
the bound on $\epsilon$ is trivial either because $\epsilon_{\rm max}\gg 1$ or
because $\epsilon_{\rm max}<0$.
However, as the typical value we find $\epsilon_{\rm max}\simeq \mFP^2/V$.
Therefore, close to the Higuchi bound the limit on the gradient
of the scalar potential is most stringent.
In other words, for $\epsilon\lesssim \mFP^2/V$ the Higuchi bound is guaranteed
to be satisfied.

\bibliographystyle{JHEP}
\bibliography{HiguchiFinal}

\end{document}